\documentclass[12pt,fleqn,a4paper,oneside]{article}
\usepackage{geometry}
\usepackage{amssymb,amsmath}
\usepackage{amsthm}
\usepackage{graphicx, float,subcaption}
\usepackage[final]{showkeys}
\usepackage{breqn}
\usepackage{adjustbox}
\theoremstyle{plain}
\newtheorem{thm}{Theorem}

\theoremstyle{definition}
\newtheorem{defn}{Definition} 

\makeatletter
\newdimen\mymathindent
{\@beginparpenalty\predisplaypenalty
	\@endparpenalty\postdisplaypenalty
	\refstepcounter{equation}%
	\trivlist \item[]\leavevmode
	\hb@xt@\linewidth\bgroup $\m@th
	\displaystyle
	\hskip\mymathindent}%
{$\hfil 
	\displaywidth\linewidth\hbox{\@eqnnum}%
	\egroup
	\endtrivlist}
\makeatother

\begin{document}
\newcommand*{\Scale}[2][4]{\scalebox{#1}{\ensuremath{#2}}}
	\newcommand{\inorm}[1]{\lVert #1\rVert_{\infty}}
	\newcommand{\mnorm}[1]{\lvert #1\rvert}
	\newcommand{\ml}[2]{E_{#1}(t^{#1} #2)}
	\newcommand{\gml}[3]{E_{#1}(t^{#2} #3)}
	\newcommand{\Gml}[3]{E_{#1}({#2}\:#3)}
	\newcommand{\pder}[2]{\dfrac{\partial^{#1}}{\partial #2^{#1}}}
	\newcommand{\norm}[1]{\lVert #1\rVert}
	\newcommand{\mll}[2]{E_{#1}(#2 t^{#1})}

\begin{center}
	\Large{Chaotic dynamics of fractional Vallis system for El-Ni\~{n}o}
	\large{Amey S. Deshpande\footnote{\noindent Department of Mathematics, IIT Bombay, Mumbai - 400076, India.} \footnote{Email: 2009asdeshpande@gmail.com, ameyd@math.iitb.ac.in.} and Varsha Daftardar-Gejji\footnote{\noindent Department of Mathematics, Savitribai Phule Pune University, Pune - 411007, India.} \footnote{Email: vsgejji@gmail.com, vsgejji@unipune.ac.in.}}
\end{center}
\begin{abstract}
Vallis proposed a simple model for El-Ni\~{n}o weather phenomenon (referred as Vallis system) by adding an additional parameter $p$ to the Lorenz system. He showed that the chaotic behavior of the Vallis system is related to the El-Ni\~{n}o effect. In the present article we study fractional version of Vallis system in depth. We investigate bifurcations and chaos present in the fractional Vallis system along with the effect of variation of system parameter $p$. It is observed that the range of values of parameter $p$ for which the Vallis system is chaotic, reduces with the reduction of the fractional order. 

Further we analyze the incommensurate fractional Vallis system and find the critical value below which the system loses chaos. We also synchronize Vallis system with Bhalekar-Gejji system.    	
\end{abstract}

\section{Introduction}
El-Ni\~{n}o  is a weather phenomenon associated with the collection of band of warm ocean water in the east-central equatorial Pacific ocean including the Pacific coast of South America. This phenomenon is a part of El-Ni\~{n}o Southern Oscillation (ENSO) climatic system which is known to have a great impact on the global climate \cite{jones2007observations}. Various mathematical models exist which describe El-Ni\~{n}o  phenomenon \cite{tung2007topics}. These models range from simplistic to highly complex. One of the simple models is proposed by Vallis in 1986 \cite{vallis1986nino} which consists of set of three autonomous non-linear differential equations (hereafter referred as Vallis system). Vallis system is a modification of Lorenz system \cite{Lorenz} with addition of a system parameter $p$. Vallis proved the existence of chaos in Vallis system and also showed that  El-Ni\~{n}o phenomenon is related to the chaotic behavior of the Vallis system \cite{JGRC:JGRC4314}.   

Over the years, integer order Vallis system has been analyzed in various articles \cite{GARAY2015253,Euzébio20143455,doi:10.1142/S021812740802121X}.   
The chaos and periodicity of Vallis model has been investigated in depth for the integer order model over $B -c$ parameter space \cite{Borghezan201715}.  

Fractional order Vallis system is not explored much. Only a couple of papers have appeared recently. Alkahtani \textit{et al.} \cite{alkahtani2016chaos} have studied Vallis model with local derivatives, Caputo derivatives and Caputo-Fabrizio derivative. They have also drawn phase portraits for commensurate fractional Vallis system with Caputo derivative for $p = 0.3$ and $\alpha = 0.85, 0.55, \: 0.25$. Merdan \cite{merdan2013numerical} have calculated numerical solution for fractional order Vallis sysem for parameter $p=0$ by  using multi-step differential transformation method. 

In the present article, fractional version of commensurate as well as incommensurate Vallis system has been studied in detail for all values of the parameter $p$. For $p \neq 0 $ this analysis becomes more tedious. Our main objective is to study the effect of fractional orders on the chaotic behavior of the Vallis system. 

The present article is organized as follows. Section \ref{6sec:prelim} introduces preliminaries and notations used in the article. Section \ref{anal} presents stability analysis of fractional order Vallis system. Section \ref{paramp} analyzes commensurate fractional Vallis system w.r.t. parameter $p$ and fractional order $\alpha$. Section \ref{incom} deals with incommensurate fractional Vallis system. Section \ref{synch} describes synchronization of Vallis system. Section \ref{conclusions} presents conclusions.
   
\section{Preliminaries and Notations} \label{6sec:prelim}
This section introduces the basic definitions and notations used throughout this paper   \cite{meiss2007differential,podlubny1999fractional,stabsurvey}.
\begin{defn}[\cite{podlubny1999fractional}]
	The Riemann-Liouville fractional integral of order $\alpha > 0$ of $f \in C^{0}$ is defined as
	\begin{align}
		I^{\alpha}f(t)=\frac{1}{\Gamma(\alpha)}\int\limits_{0}^{t}\frac{f(\tau)}{(t-\tau)^{1-\alpha}}d\tau.
	\end{align}
\end{defn}

\begin{defn}[\cite{podlubny1999fractional}]
	The Caputo derivative of order $\alpha \in (k-1, k], ~ k \in \mathbb{N}$ of $f \in C^k$ is defined as 
	\begin{align}
		\begin{split}
			D^{\alpha}f(t) &= I^{k-\alpha}f^{(k)}(t) \\&= \frac{1}{\Gamma(k-\alpha)} ~\int_{0}^{t} (t-\tau)^{k-\alpha-1} f^{(k)}(\tau)\: d\tau,~~ \alpha \in (k-1,k), 
		\end{split}\\
		D^{\alpha}f(t) &= f^{(k)}(t), ~ \alpha = k.
	\end{align}
\end{defn}

\textbf{Hereafter in the article, we always assume Caputo derivative of order $ \alpha, \, 0 < \alpha \leq 1$, unless stated otherwise.} 

\begin{defn}\cite{stabsurvey}
	For $0 < \alpha_i \leq 1,$ and $ f_i \in C^{1},~(i=1,2,3)$, fractional dynamical system along with initial conditions is given as 
	\begin{align} \label{pb_def1}
	\begin{split}
		D^{\alpha_1} x(t) &= f_1(x(t),y(t),z(t)),\\
		D^{\alpha_2} y(t) &= f_2(x(t),y(t),z(t)),\\
		D^{\alpha_3} z(t) &= f_3(x(t),y(t),z(t)),\\
		(x(0), y(0),z(0)) &= (x_0,y_0,z_0).
		\end{split}
	\end{align}
\end{defn}

If $\alpha_1 = \alpha_2 = \alpha_3$, then the system \eqref{pb_def1} is called as \textit{commensurate} system and this common fractional order is denoted as $\alpha$. System is called as \textit{incommensurate} if it is not commensurate\cite{tavazoei2008chaotic}. For $\alpha_1 = \alpha_2 = \alpha_3 = 1,$ the system \eqref{pb_def1} reduces to integer order dynamical system.

Denote $\Sigma = \alpha_1 + \alpha_2 + \alpha_3$.

\begin{defn}
	The critical value ($\Sigma_{cr}$) is defined as the largest value of $\Sigma$, such that for the $\Sigma \leq \Sigma_{cr}$, the system \eqref{pb_def1} is not chaotic. 
\end{defn}

\begin{defn}
	The point $(x^*,y^*,z^*) \in \mathbb{R}^3$ is called as equilibrium point of the system \eqref{pb_def1}, if $f_i(x^*,y^*,z^*) = 0, ~~ i=1,2,3$. 
\end{defn}

\subsection{Stability analysis of the commensurate fractional systems}
Consider the system \eqref{pb_def1}, with $\alpha_1 = \alpha_2 = \alpha_3 \in (0,1]$. Let $f=(f_1,f_2,f_3)$ and $X=(x,y,z), X^* = (x^*,y^*,z^*) \in \mathbb{R}^3$.

An equilibrium point $X^*$ of the system \eqref{pb_def1} is called as a hyperbolic equilibrium point if $\mnorm{\arg(\lambda)} \neq \frac{\pi \alpha}{2}$, for every eigenvalue $\lambda$ of the matrix $J = Df(X^{*})$ \cite{matignon1996stability}. Assume $X^{*}$ is an  equilibrium point of the system \eqref{pb_def1}. Let $\xi = X-X^{*} \in \mathbb{R}^3$, then \cite{stabsurvey}
\begin{align*}
	D^{\alpha} \xi &= D^{\alpha} (X - X^{*})\\
	&=D^{\alpha} X\\
	& = f(\xi + X^{*})\\
	&= f(X^{*}) + Df(X^{*}) \xi  + \cdots 
\end{align*}
Thus we get,
\begin{equation} \label{6eq:stab1}
	D^{\alpha} \xi = J \xi, ~~\alpha = (\alpha_1, \alpha_2, \alpha_3),~\alpha_1 = \alpha_2 = \alpha_3 \in (0,1].
\end{equation}

The eigenvalues of the matrix $J$ determine the stability properties of the system about an equilibrium point $X^{*}$ \cite{stabsurvey}. The stability criteria is due to Matignon \cite{matignon1996stability} which states: If $X^{*}$ is a hyperbolic equilibrium point then the the trajectories of the system \eqref{6eq:stab1} are asymptotically stable if and only if $\mnorm{\arg(\lambda)} > \frac{\alpha \pi}{2}$, for every eigenvalue $\lambda$  of $J$.
\subsection{Numerical Method}\label{6sec:NumericalMethods}
Various  numerical methods exist in the literature for solving nonlinear fractional differential equations. For simulations pertaining to fractional order dynamical systems, predictor-corrector type methods such as Fractional Adams Method (FAM) or New Predictor Corrector Method (NPCM) are more suitable and are extensively used in the literature. 

In the present article, we have used New Predictor Corrector Method (NPCM) which is proposed by Daftardar-Gejji \textit{et al.} \cite{daftardar2014new}. This method is more time efficient, accurate and has better stability properties. 

We briefly describe NPCM below. 
Consider the following fractional initial value problem:
\begin{equation}\label{1eq:fam1}
	D^{\alpha}y(t) = f(t,y(t)),~ \frac{d^k y(0)}{dt^k}  = y_0^{k}, ~ k=0,1,\cdots,m-1,
\end{equation} where $D^{\alpha}$ denotes Caputo derivative, with order $\alpha, m-1 < \alpha < m$, $m \in \mathbb{N}$.
Then eqn. \eqref{1eq:fam1} is equivalent to the integral equation
\begin{equation}\label{1eq:fam2}
	y(t) = \sum_{k=0}^{m-1} \frac{t^k}{k!} y_0^{k} + \frac{1}{\Gamma(\alpha)} \int_{0}^{t} (t-\tau)^{\alpha-1} f(\tau, y(\tau))~d\tau.
\end{equation}
In this method we discretize integral eqn.  \eqref{1eq:fam2} on a uniform grid $t_j = j h, ~ j=0,1,\cdots,N$, $h= \frac{T}{N}$, on $[0,T]$. Assume $y_1, y_2, \cdots, y_k$, values at first $k$ points, are known. To find $y_{k+1}$, we define predictors
\begin{align} \label{1eq:npcm2}
	y_{k+1}^{P} &= \sum_{j=0}^{m-1} \frac{t_{k+1}^j}{j!} y_0^{j} + \frac{h^{\alpha}}{\Gamma(\alpha + 2)} \sum_{j= 0}^{k} a_{j, \: k+1} f(t_j, y_j),\\ \label{1eq:npcm3}
	z_{k+1}^{P} &= \frac{h^{\alpha}}{\Gamma(\alpha + 2)} f(t_{k+1}, y_{k+1}^{P}), 
\end{align} where $a_{j, \: k+1}$ are defined as
\begin{equation}\label{1eq:fama}
	a_{j, \:k+1} = \begin{cases}
		k^{\alpha + 1} - (k-\alpha)(k+1)^{\alpha},~~~~~~~~~~j=0,\\
		(k-j+2)^{\alpha +1} + (k-j)^{\alpha + 1} - 2 (k-j+1)^{\alpha + 1}, ~~ 1 \leq j \leq k, \\
		1, ~~~~~~~~~~j=k+1. 
	\end{cases}
\end{equation}
Using predicted values, we get the corrected value of $y_{k+1}$ as 
\begin{equation}\label{1eq:npcm1}
	y_{k+1} = y_{k+1}^{P} + \frac{h^{\alpha}}{\Gamma(\alpha + 2)} f(t_{k+1}, y_{k+1}^{P} + z_{k+1}^{P}).
\end{equation}

Equations eqn. \eqref{1eq:npcm2} to  eqn. \eqref{1eq:npcm1}, constitute the new predictor corrector method \cite{daftardar2014new}. 

\begin{thm}[Error estimate \cite{daftardar2014new}]
	Assume that $D^{\alpha} y$ is $C^2$ over interval $[0,T]$ and $y(t_j)$ denote exact values of $y$ at point $t_j$ then  
	\begin{equation}\label{1eq:npcmerror}
		\max_{0 \leq j \leq N} \mnorm{y(t_j) - y_j} = 	O(h^2), ~~ \alpha > 0.
	\end{equation} 
\end{thm}

\section{Analysis of fractional Order Vallis system} \label{anal}

The fractional version of Vallis system is given below.

\begin{align}\label{vallis1}
\begin{split}
D^{\alpha_1} x &= B y - C (x + p)\\
D^{\alpha_2} y &= x z - y \\
D^{\alpha_3} z &= -x y - z + 1 
\end{split}
\end{align}
where $0 < \alpha_i \leq 1, ~ i=1,2,3$. The system \eqref{vallis1}, is called as commensurate if $\alpha_1 = \alpha_2 = \alpha_3 $ and incommensurate otherwise. The system reduces to fractional Lorenz system whenever $p=0,\: B=1$.

This model is developed by treating equatorial ocean as a box of fluid characterized by temperatures in the east and west. Variable $x$ represents current generated by the temperature gradient, co-ordinate $y$ represents half of the difference of east-west temperatures and $z$ represents the average of the east and west temperatures. The parameter $B$ governs strength of air-sea interactions and the vertical temperature difference, while parameter $C$ represents the ratio of time scales of decay of sea-surface temperature anomalies to a frictional time scale. The parameter $p$ measures the average effect of equatorial winds on the sea 
\cite{JGRC:JGRC4314,Borghezan201715}.

\begin{figure}[h]
	\centering
	\includegraphics[width=0.7\linewidth]{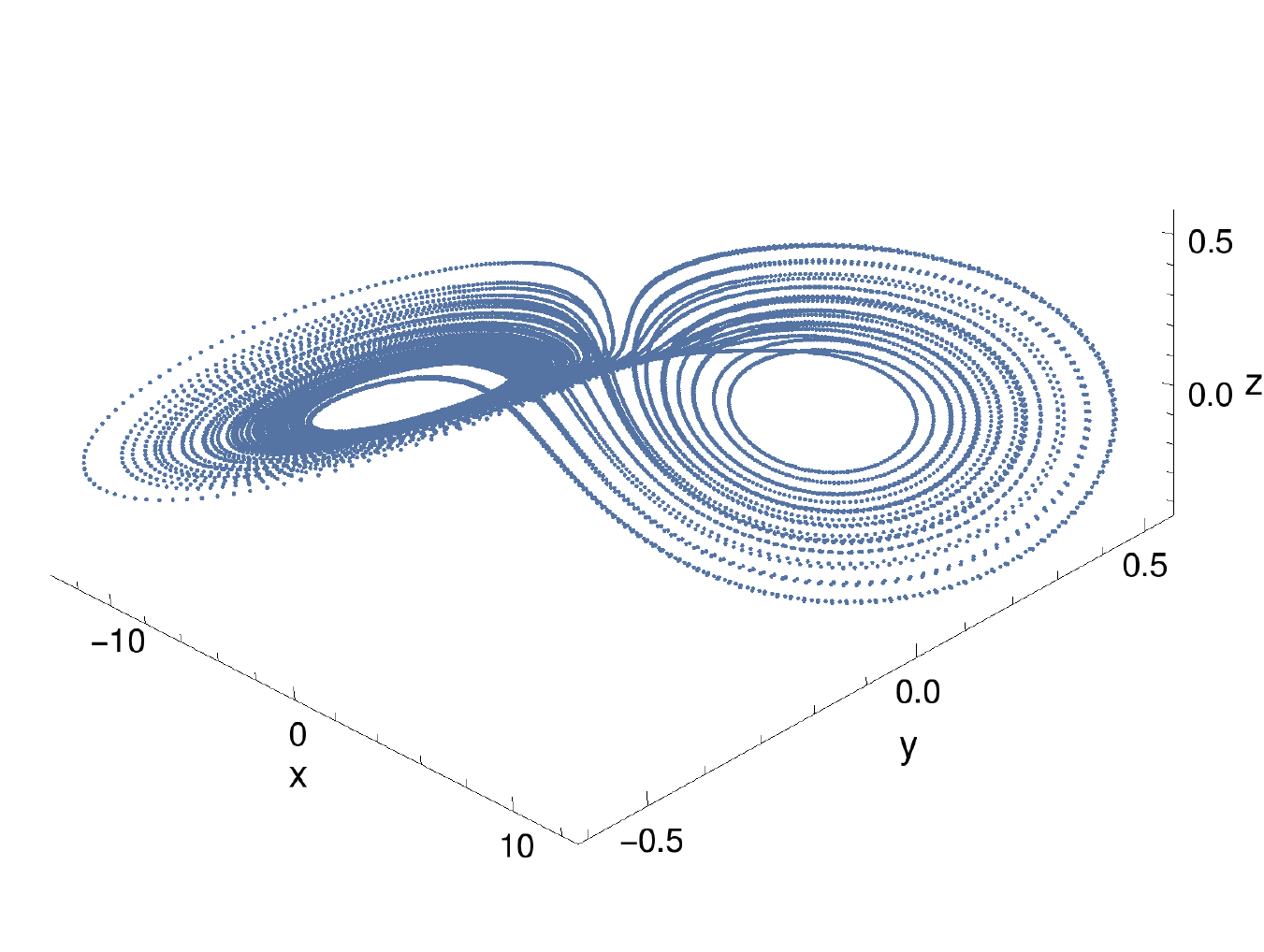}
	\caption{Chaotic attractor of system \eqref{vallis1} for $\alpha=0.98$, and $p=0.35,\: B=150,\: C=4$.}
	\label{fig:vallis98-35}
\end{figure}

The equilibrium points of the system \eqref{vallis1} are $F_1\equiv(x_1,y_1,z_1)$, $F_2\equiv(x_2,y_2,z_2)$, $F_3\equiv(x_3,y_3,z_3)$, where $x_1, x_2, x_3$ are roots of the polynomial
\begin{equation} \label{poly1}
C \xi^3 + C p \xi^2 + (C-B) \xi + C p, 
\end{equation}
and $y_i$, $z_i$ are given as
\begin{equation}
y_i = \frac{x_i}{1 + x_i^2}, ~~z_i = \frac{1}{1 + x_i^2},~~i=1,2,3.
\end{equation}  
The Jacobian matrix $J$ and characteristic polynomial for $J$ are given as
\begin{align}\label{jacob}
J &= \begin{pmatrix}
-C & B & 0 \\ 
z & -1 & x \\ 
-x & -y & -1
\end{pmatrix},\\ \label{character}
\phi(\lambda) &= \lambda^3 + (2+ C) \lambda^2 + (1+2 C + x y - B z) \lambda + (C + B x^2 + C x y - B z). 
\end{align} 
For the values $B=150$, $p=0.35$ and $C=4$, the equilibrium points are $F_1 \equiv (0.0095,0.0095,0.9999)$,  $F_2 \equiv (5.86412,0.16571,0.02825)$ and \\$F_3\equiv (-6.22371, -0.156632, 0.025167)$. The corresponding eigenvalues of the Jacobian matrix are given in Table (\ref{tab1}). Further the phase portrait of commensurate system \eqref{vallis1} with fractional order $\alpha  = 0.98$   is drawn in fig. (\ref{fig:vallis98-35}). 
A positive value of the largest Lyapunov exponent (LLE) indicates a chaotic behavior. In this case, the LLE of system \eqref{vallis1}, calculated by using Wolf algorithm \cite{WOLF1985285}, is found out to be LLE = $0.4181$. 

\begin{table}[H]
	\centering
\begin{tabular}{|c|c|c|c|}
	\hline 
	& $\lambda_1$ & $\lambda_2$ & $\lambda_3$ \\ 
\hline
	$F_1$ & $-14.8384$ & $-0.9999$ & $9.83835$ \\ 

	$F_2$ & $-19.4082$ & $6.704-14.866 i$ & $6.704+14.866 i$ \\ 

	$F_3$ & $-20.093$ & $7.0464-15.4828 i$ & $7.0464+15.4828 i$ \\ 
	\hline 
\end{tabular} 
\caption{Eigenvalues of Jacobian matrix $\lambda_1, \lambda_2, \lambda_3$, around equilibrium points $F_1, F_2, F_3$.}
\label{tab1}
\end{table}

\section{Commensurate Vallis System} \label{paramp}
\subsection{Chaos and Bifurcation w.r.t. parameter $p$}
 We analyze system \eqref{vallis1} by varying the system parameter $p$.
 Let $B=150, C=4$. We analyze stability of the system \eqref{vallis1} by varying parameter $p$. The eqn. \eqref{poly1} has at least one real root. Using discriminant $\Delta$ of the cubic polynomial we see that eqn. \eqref{poly1} will have all three real roots whenever $\mnorm{p} \leq 17.7754$. 
 Thus we see that for $\mnorm{p}=17.7754$ system \eqref{vallis1} undergoes a pitchfork bifurcation.

Using eqn. \eqref{character}, we numerically calculate eigenvalues around each equilibrium point.
From these calculations we observe that, for $\mnorm{p} \leq 17.35$, equilibrium point $F_1$ is index 1 saddle with all three real eigenvalues while $F_2, \:F_3$ are index 2 saddles with one real and two complex conjugate eigenvalues with positive real part. Thus for $\mnorm{p} \leq 17.35$ system \eqref{vallis1} satisfies necessary condition for the existence of chaos. 
Note that for fractional order $0 < \alpha \leq 1$, the values of $p$ for which system may be chaotic is given as $\mnorm{p} < M$ where $0 < M \leq 17.35$. This is due to the stability criteria of fractional systems \cite{matignon1996stability}. 

\begin{figure}[H]
		\centering
	\includegraphics[width=0.6\linewidth]{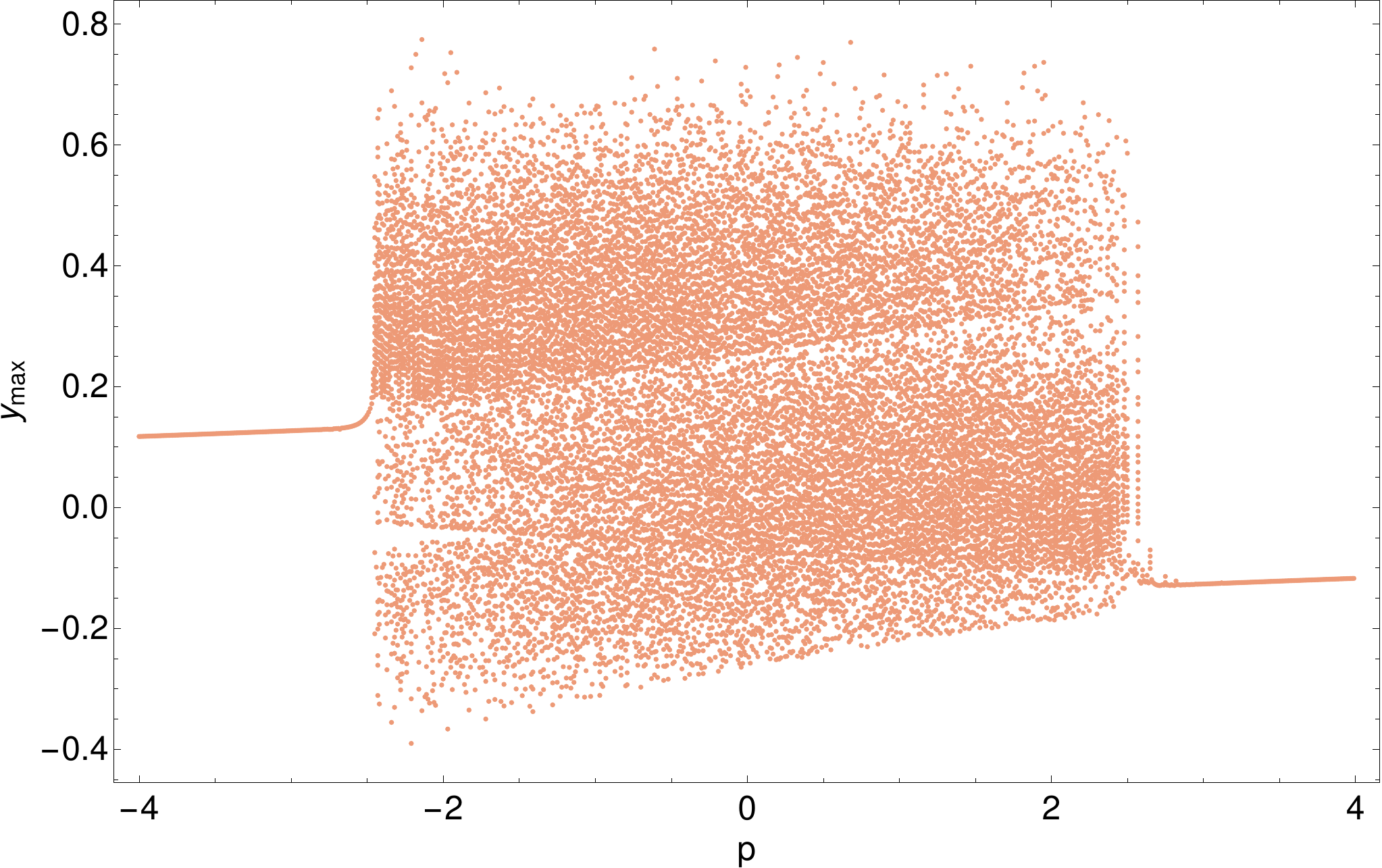}
	\caption{Bifurcation diagram versus $p$ for $\alpha = 1$}
	\label{fig:1-p-bif}
\end{figure}

\begin{figure}[H]
	\centering
	\begin{subfigure}{0.45\linewidth}
	\includegraphics[width=\linewidth]{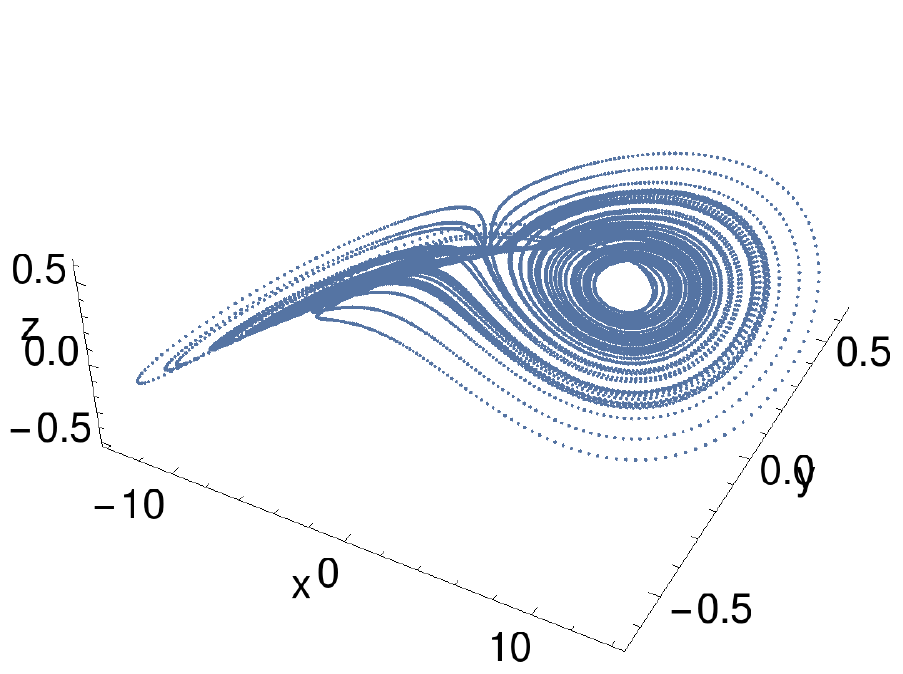}
	\end{subfigure}
	\begin{subfigure}{0.45\linewidth}
	\includegraphics[width=\linewidth]{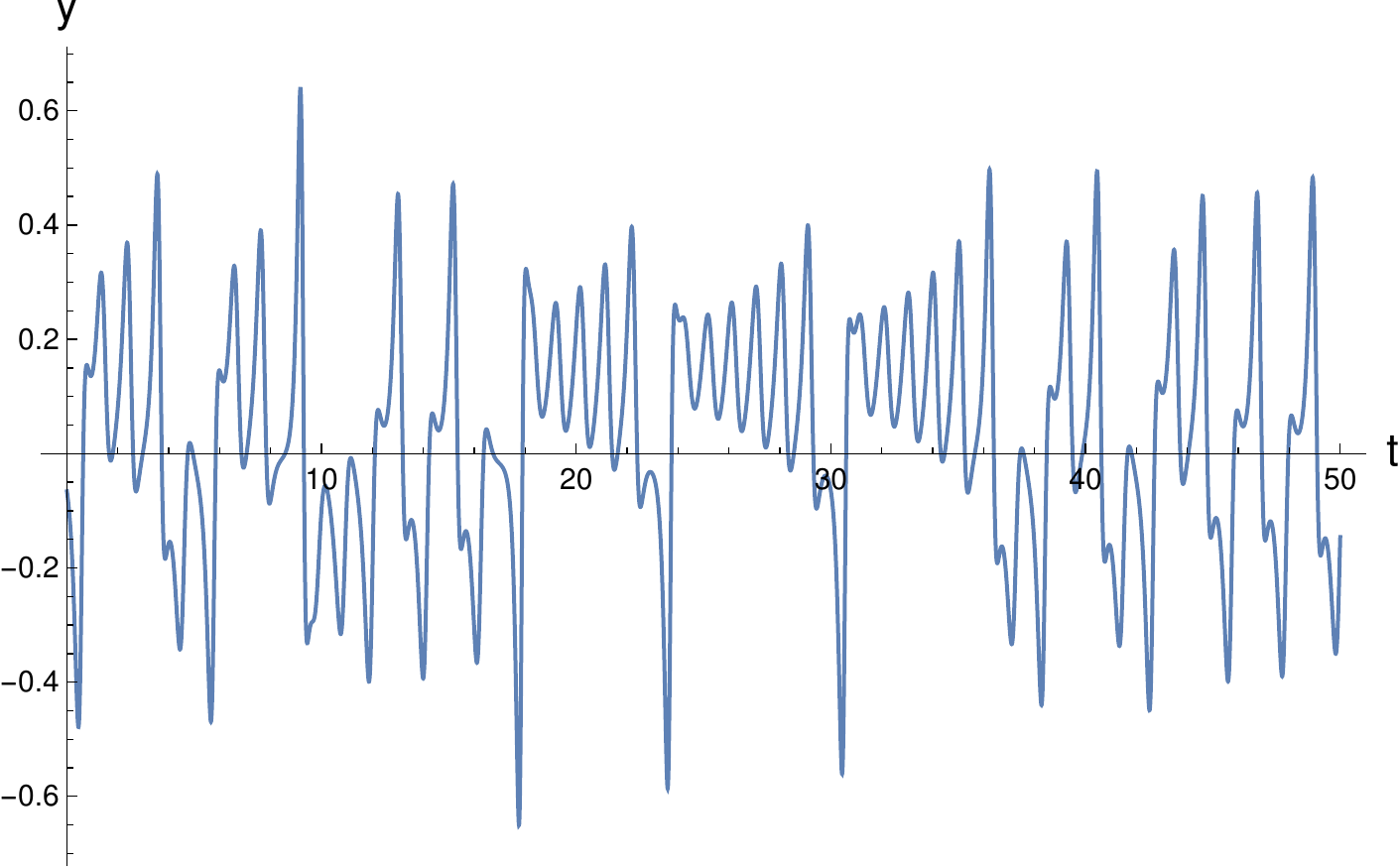}
\end{subfigure}
	\caption{Phase and time-series for $p=-0.5$ for $\alpha = 1$}
	\label{fig:1-p-05}
\end{figure}

The bifurcation diagram versus parameter $p$ for fractional order $\alpha = 1$ (\textit{c.f.} fig. (\ref{fig:1-p-bif})) shows that for $-2.45 \leq p \leq 2.45$, system is chaotic. The system will break into El-Ni\~{n}o effect, if the difference between the eastern and western sea temperatures is high. Fig. (\ref{fig:1-p-05}) shows that $p=-0.5$, the chances of system breaking into El-Ni\~{n}o are higher than for $p=0$, while when $\mnorm{p} > 2.45$, system stabilizes to one of the equilibrium points and hence does not admit El-Ni\~{n}o effect.

\begin{figure}[H]
	\centering
	\begin{subfigure}{0.8\linewidth}
		\includegraphics[height=0.33\textheight]{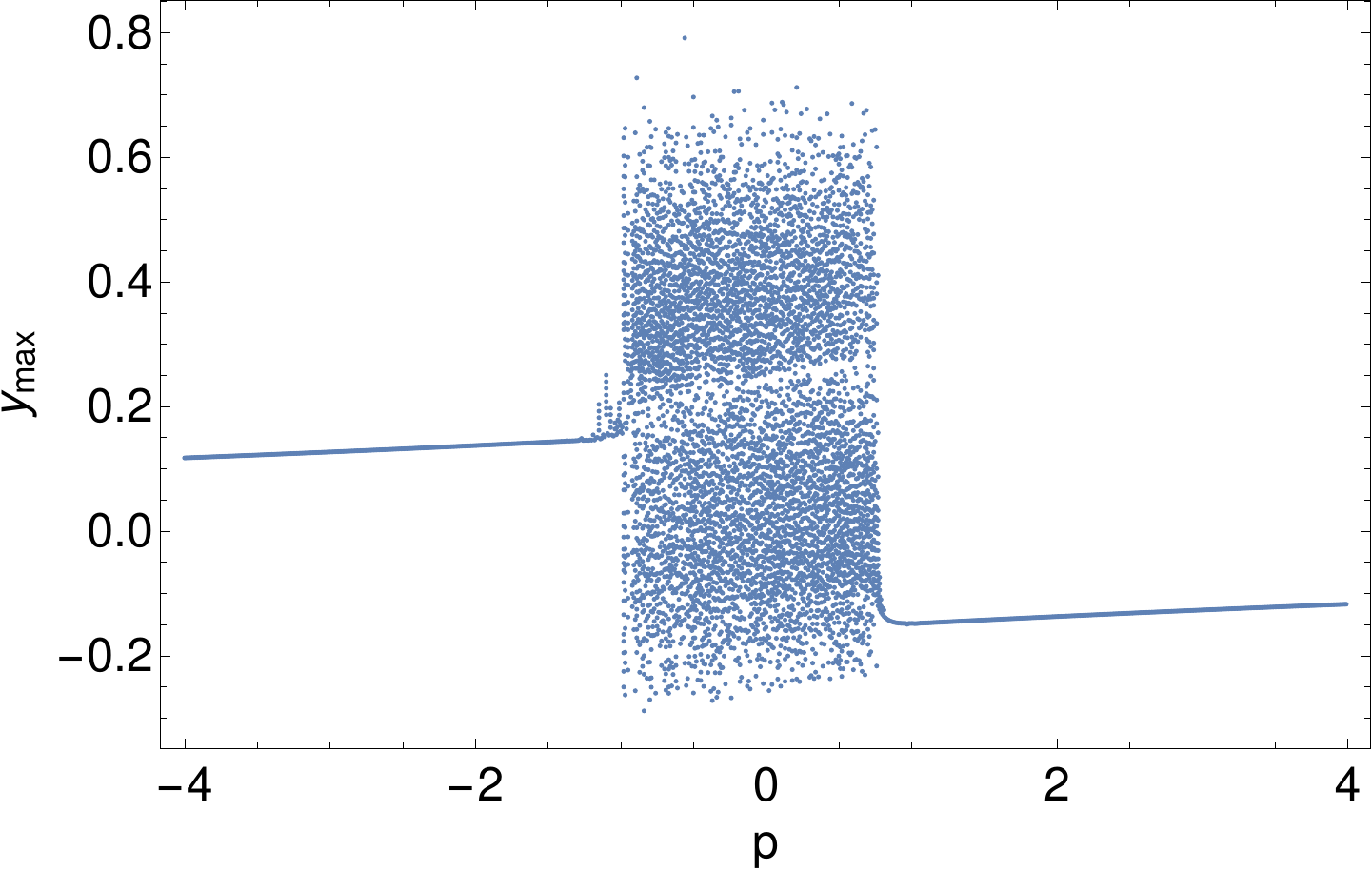}
	\end{subfigure}\\
	\begin{subfigure}{0.8\linewidth}
		\includegraphics[height=0.33\textheight]{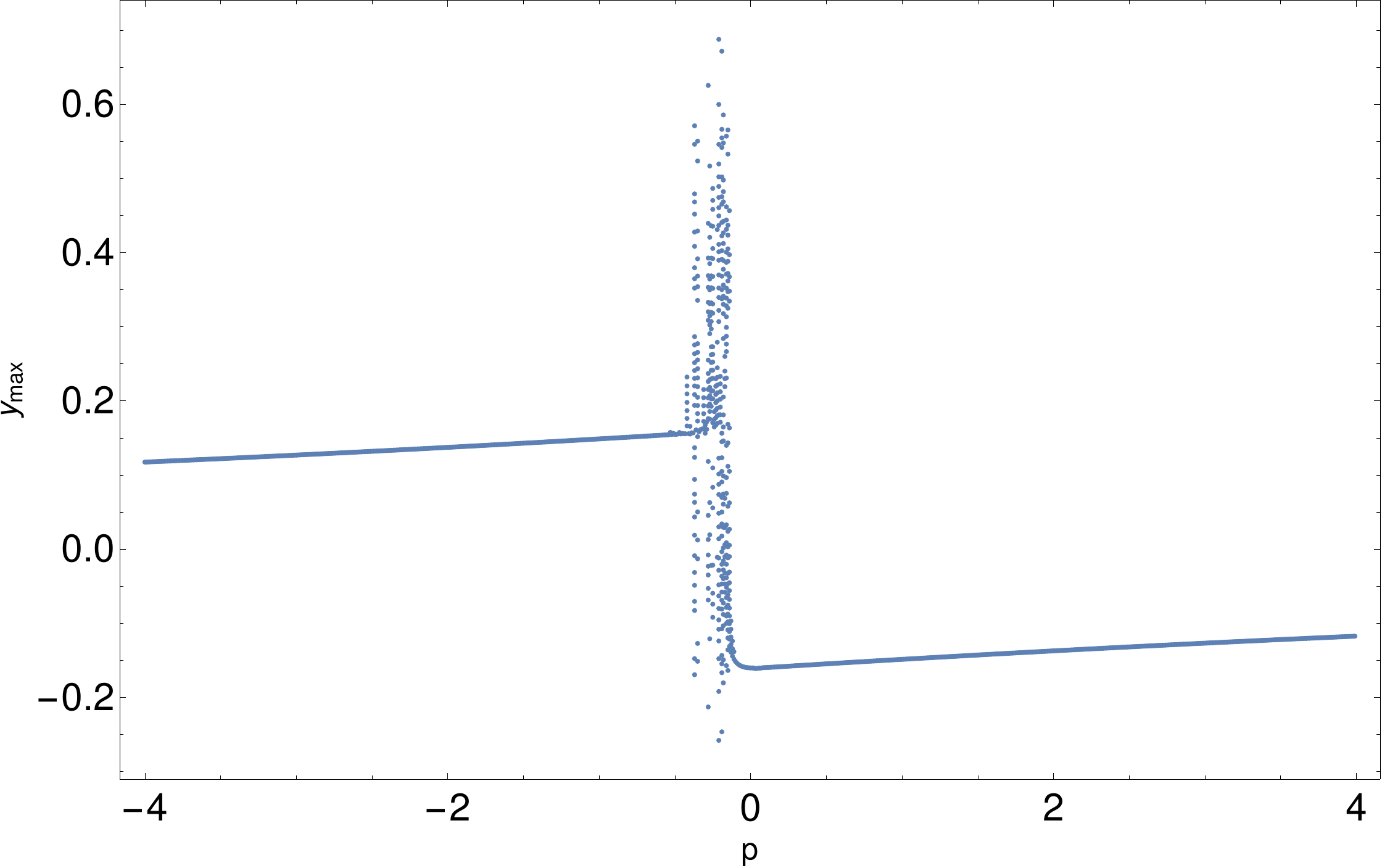}
	\end{subfigure}\\
	\begin{subfigure}{0.8\linewidth}
	\includegraphics[height=0.33\textheight]{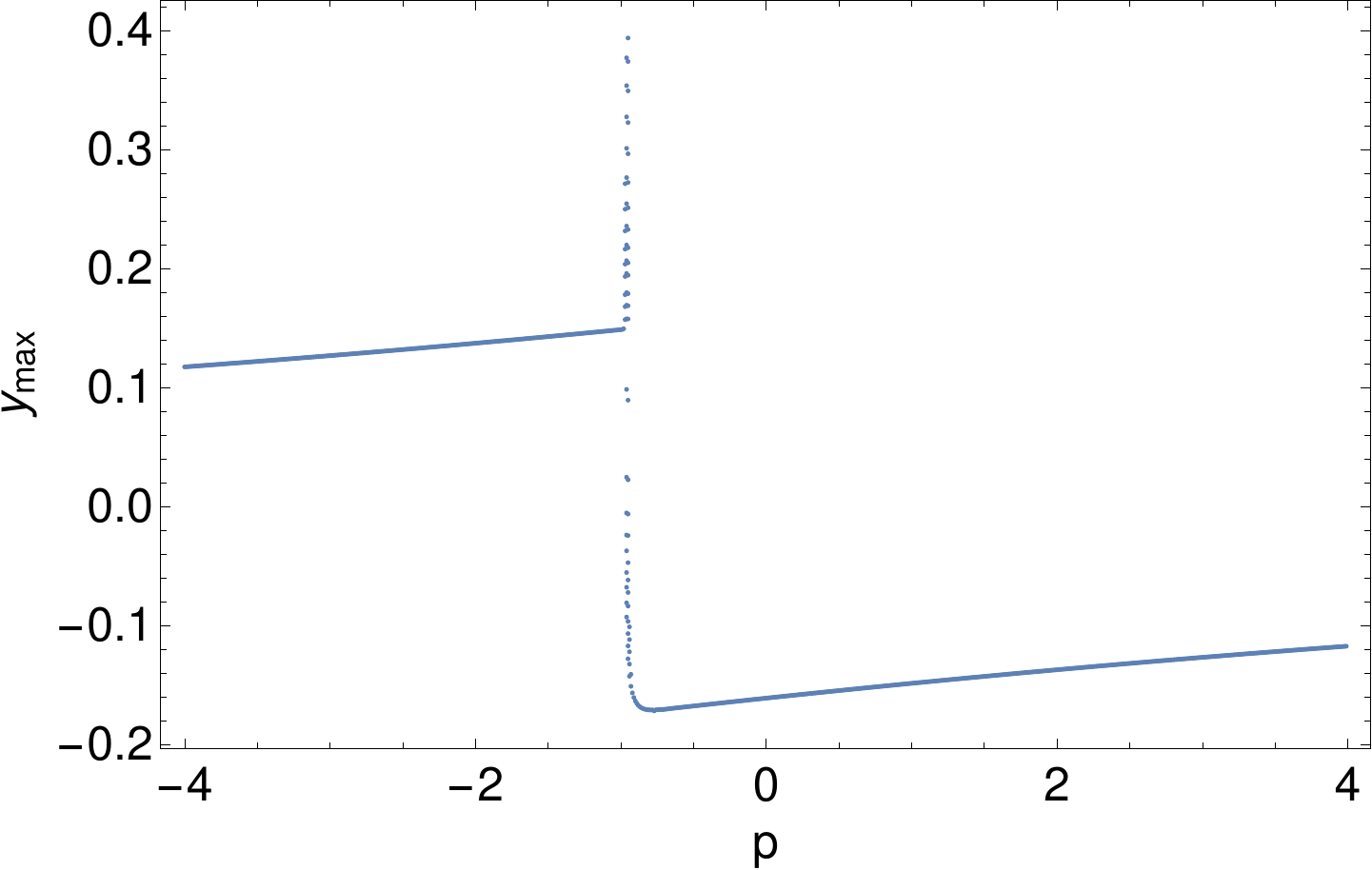}
\end{subfigure}
	\caption{Bifurcation diagrams versus parameter $p$ for fractional orders (a) $\alpha = 0.985$ (b) $\alpha = 0.975$ (c) $\alpha = 0.965$ }
	\label{fig:frac-p-bif}
\end{figure}

The bifurcation diagrams w.r.t. parameter $p$ corresponding to orders $\alpha = 0.985, \:0.975,\: 0.965$ are shown in figure (\ref{fig:frac-p-bif}). For $\alpha = 0.985$, the chaotic region shrinks to $-1.5 \leq p \leq 1.5$, while for $\alpha = 0.975$, region further shrinks to $-0.5 \leq p < 0$. This reduction in chaotic region reduces the chances of system breaking into El-Ni\~{n}o. For $\alpha = 0.965$, system \eqref{vallis1} completely loses chaos for all values of $p$.  

Table (\ref{Tab:p}) shows the calculated values of LLE for various values of $p$ and $\alpha = 1, 0.985, 0.975, 0.965$. A positive value of LLE indicates existence of chaos while negative or zero value confirms non-existence of chaotic behavior.

\begin{table}[H]
\centering
\begin{tabular}{|c|cccc|}
	\hline 
	$p$ & $\alpha = 1$ & $\alpha = 0.985$ & $\alpha =0.975$ & $\alpha = 0.965$ \\ 
	\hline 
	$-3.5$ & $-0.1002$ & $-0.0229$ & $-0.0142$ & $-0.00072$ \\ 
	$-2.4$ & $0.2656$ & $0.00018$ & $-0.0160$ & $-0.00010$ \\ 
	$-1.5$ & $0.5115$ & $-0.0694$ & $-0.015$ & $-0.0196$ \\ 
	$-0.5$ & $0.4735$ & $0.4970$ & $-0.076$ & $-0.06522$ \\ 
	$0$ & $0.5069$ & $0.49427$ & $-0.0055$ & $-0.9$ \\ 
	$0.35$ & $0.52828$ & $0.2133$ & $-0.010$ & $-0.9$ \\ 
	$1.5$ & $0.5350$ & $0.0793$ & $-0.9$ & $-0.0332$ \\ 
	$2.4$ & $0.2317$ & $0.04$ & $-0.9$ & $-0.0245$ \\ 
	$3.5$ & $-0.1270$ & $-0.032$ & $-0.9$ & $-0.0168$ \\ 
	\hline 
\end{tabular} 
\caption{Largest Lyapunov exponent (LLE) values of system \eqref{vallis1} for various values of parameter $p$ and fractional orders $\alpha = 1, \: 0.985, \: 0.975$ and $\alpha = 0.965$.}
\label{Tab:p}
\end{table}
\subsection{Chaos and Bifurcation w.r.t. fractional order $\alpha$ } \label{fracorder}

\begin{figure}[H]
	\centering
	\includegraphics[width=0.8\linewidth]{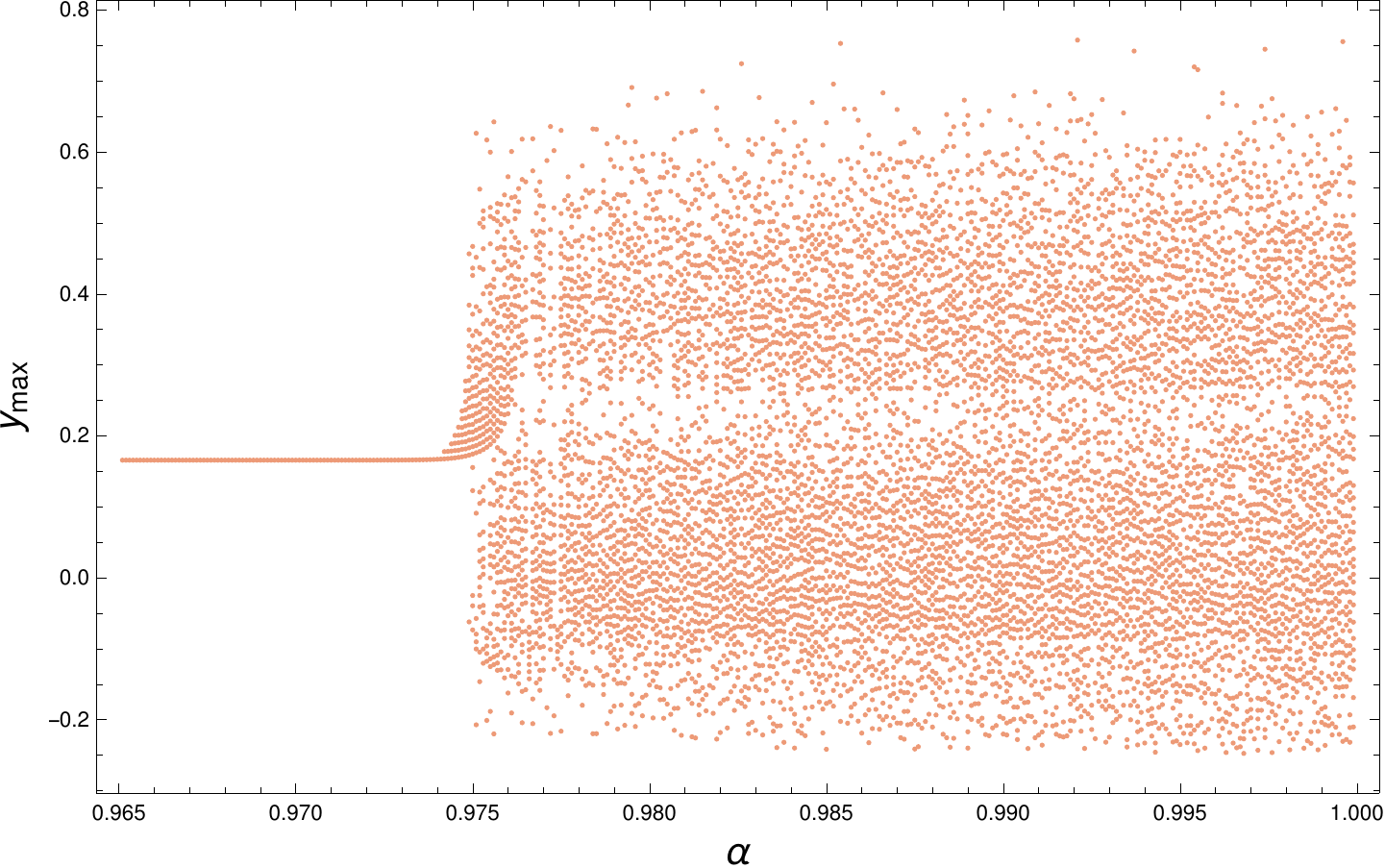}
	\caption{Bifurcation diagram versus fractional order $\alpha$}
	\label{fig:order-p35-bif}
\end{figure}

Let $B = 105$, $C = 4$ and $p=0.35$.
We analyze eqn. \eqref{vallis1}, in particular for the case $\alpha = \alpha_1 = \alpha_2 = \alpha_3$. Figure (\ref{fig:order-p35-bif}) shows bifurcation of system \eqref{vallis1} versus parameter $\alpha$. 
Table (\ref{Tab:commensurate}) shows LLE values for some values of $\alpha$.

\begin{table}[H]
	\centering
	\begin{tabular}{|c|c|}
		\hline 
		$\alpha$ & LLE \\ 
		\hline 
		$0.970$ & $-0.0246$ \\ 
		$0.976$ & $0.31576$ \\ 
		$0.985$ & $0.47899$ \\ 
		$1.000$ & $0.3683$ \\ 
		\hline 
	\end{tabular} 

	\caption{Largest Lyapunov exponents (LLE) of commensurate Vallis system for various fractional orders $\alpha$.}
	\label{Tab:commensurate}
\end{table}

From fig. (\ref{fig:order-p35-bif}) we observe that the integer order system is chaotic. As the fractional order reduces, chaotic nature of the system remains till $\alpha \approx 0.975$, below which chaos disappears. Thus the lowest order for which the commensurate system \eqref{vallis1} shows chaos is $\alpha =  0.975$.  

\section{Incommensurate Vallis System} \label{incom}
\begin{figure}[H]
 	\centering
 	\includegraphics[width=0.8\linewidth]{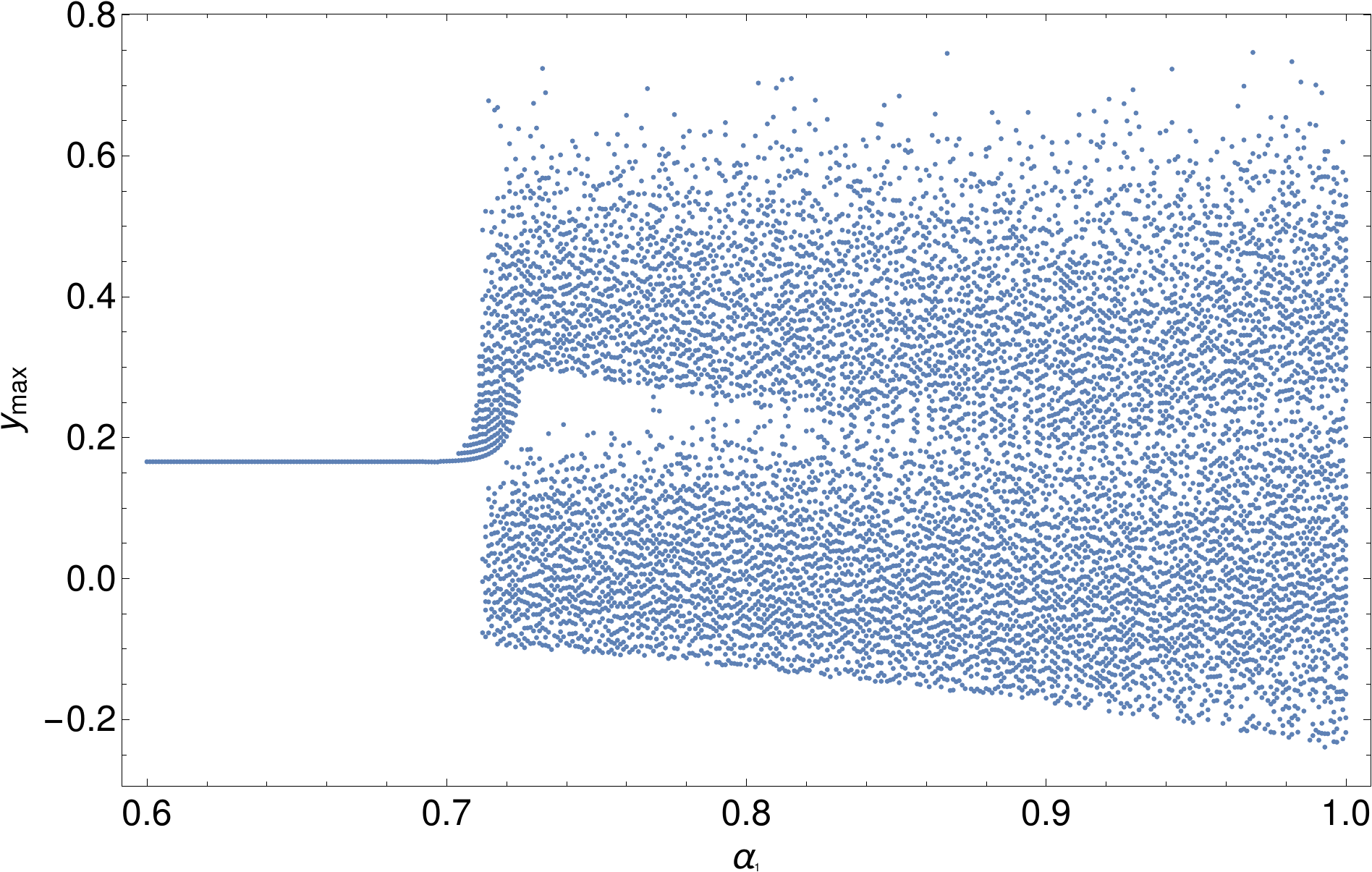}
 	\caption{Bifurcation diagram versus fractional order $\alpha_1$, $\alpha_2 = \alpha_3 = 1$}
 	\label{fig:order-incom-alpha-bif}
 \end{figure}

Consider incommensurate order system \eqref{vallis1}, for parameter values $B = 105$, $C = 4$ and $p=0.35$. We consider $\alpha_2 = \alpha_3 = 1$.  The bifurcation diagram w.r.t. $\alpha_1$ has been plotted in fig. (\ref{fig:order-incom-alpha-bif}).
Table (\ref{Tab:incom1}) shows values of LLE for some values of fractional order $\alpha_1$.

\begin{table}[H]
	\centering
	\begin{tabular}{|c|c|}
		\hline 
		$\alpha_1$ & LLE \\ 
		\hline 
		$0.670$ & $-0.00386$ \\ 
		$0.720$ & $0.2113$ \\ 
		$0.850$ & $0.4913$ \\ 
		$1.000$ & $0.41214$ \\ 
		\hline 
	\end{tabular} 

	\caption{Largest Lyapunov exponents (LLE) of incommensurate Vallis system for various values of $\alpha_1$, $\alpha_2 = \alpha_3 = 1$.}
		\label{Tab:incom1}
\end{table}

The fractional incommensurate system \eqref{vallis1} remains chaotic until $\alpha_1 = 0.71$ and below this value the system stabilizes. Thus critical order below which the system loses chaos,  is $2.71$.

 \begin{figure}[H]
	\centering
	\includegraphics[width=0.8\linewidth]{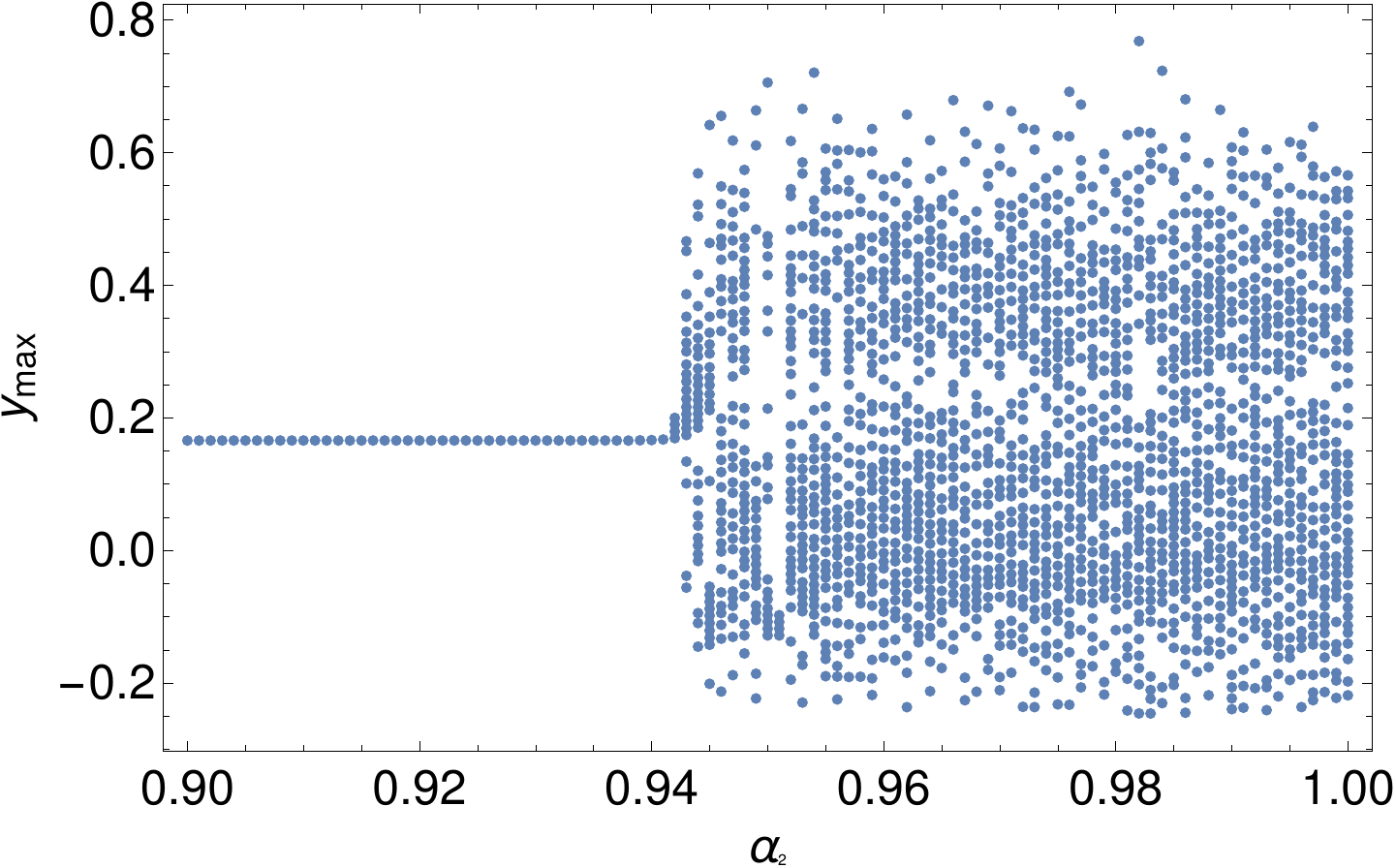}
	\caption{Bifurcation diagram versus fractional order $\alpha_2$, $\alpha_1 = \alpha_3 = 1$}
	\label{fig:order-incom-beta-bif}
\end{figure}

\begin{table}[H]
	\centering
	\begin{tabular}{|c|c|}
		\hline 
		$\alpha_2$ & LLE \\ 
		\hline 
		$0.930$ & $-0.13101$ \\ 
		$0.960$ & $0.4852$ \\ 
		$0.980$ & $0.54821$ \\ 
		$1.000$ & $0.412$ \\ 
		\hline 
	\end{tabular} 

	\caption{Largest Lyapunov exponents (LLE) of incommensurate Vallis system for various values of $\alpha_2$, $\alpha_1 = \alpha_3 = 1$.}
		\label{Tab:incom2}
\end{table}

Fig. (\ref{fig:order-incom-beta-bif}) shows bifurcation diagram versus order $\alpha_2$ (with $\alpha_1 = \alpha_3 = 1$). Further Table (\ref{Tab:incom2}) shows values of LLE for some values of fractional order $\alpha_2$. The chaos in this case, disappears below $\alpha_2 = 0.942$. Thus effective critical order for the system \eqref{vallis1} is $\Sigma = 2.942$.

 \begin{figure}[H]
	\centering
	\includegraphics[width=0.8\linewidth]{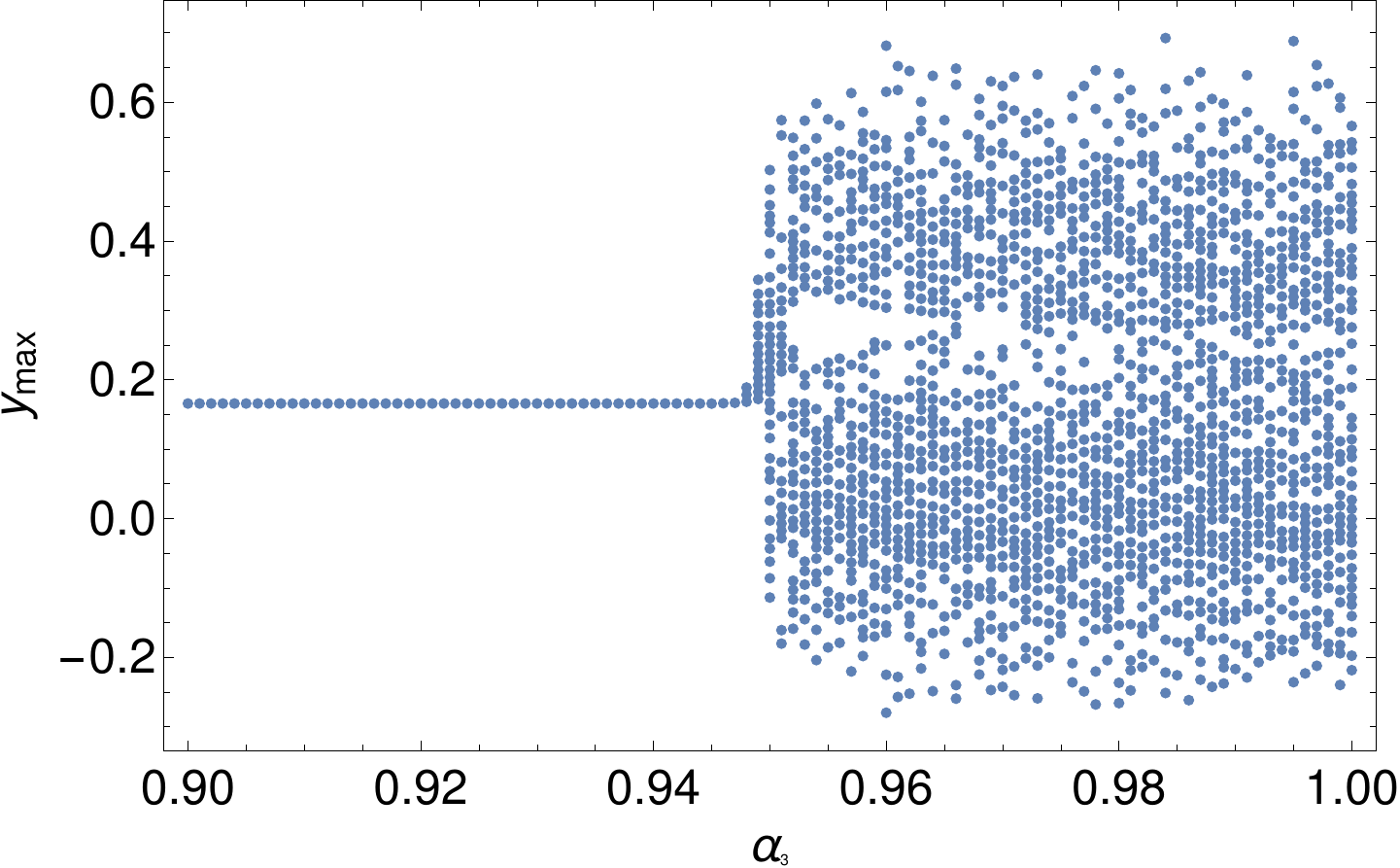}
	\caption{Bifurcation diagram versus fractional order $\alpha_3$, $\alpha_1 = \alpha_2 = 1$}
	\label{fig:order-incom-gamma-bif}
\end{figure}

\begin{table}[H]
	\centering
	\begin{tabular}{|c|c|}
		\hline 
		$\alpha_3$ & LLE \\ 
		\hline 
		$0.970$ & $-0.0246$ \\ 
		$0.976$ & $0.31576$ \\ 
		$0.985$ & $0.47899$ \\ 
		$1.000$ & $0.3683$ \\ 
		\hline 
	\end{tabular} 

	\caption{Largest Lyapunov exponents (LLE) of incommensurate Vallis system for various values of $\alpha_3$, $\alpha_1 = \alpha_2 = 1$.}
		\label{Tab:incom3}
\end{table}

Let $\alpha_1 = \alpha_2 = 1$. The bifurcation with respect to the order $\alpha_3$ is plotted in fig. (\ref{fig:order-incom-gamma-bif}). Further Table (\ref{Tab:incom3}) shows the values of LLE for some values of fractional order $\alpha_3$. It is clear that the minimum order for existence of chaos is $\alpha_3 = 0.948$. Hence the critical order is $\Sigma = 2.948$.   

\section{Synchronization of Vallis system and Bhalekar-Gejji system}\label{synch}
\begin{figure}[H]
	\centering
	\includegraphics[width=0.8\linewidth]{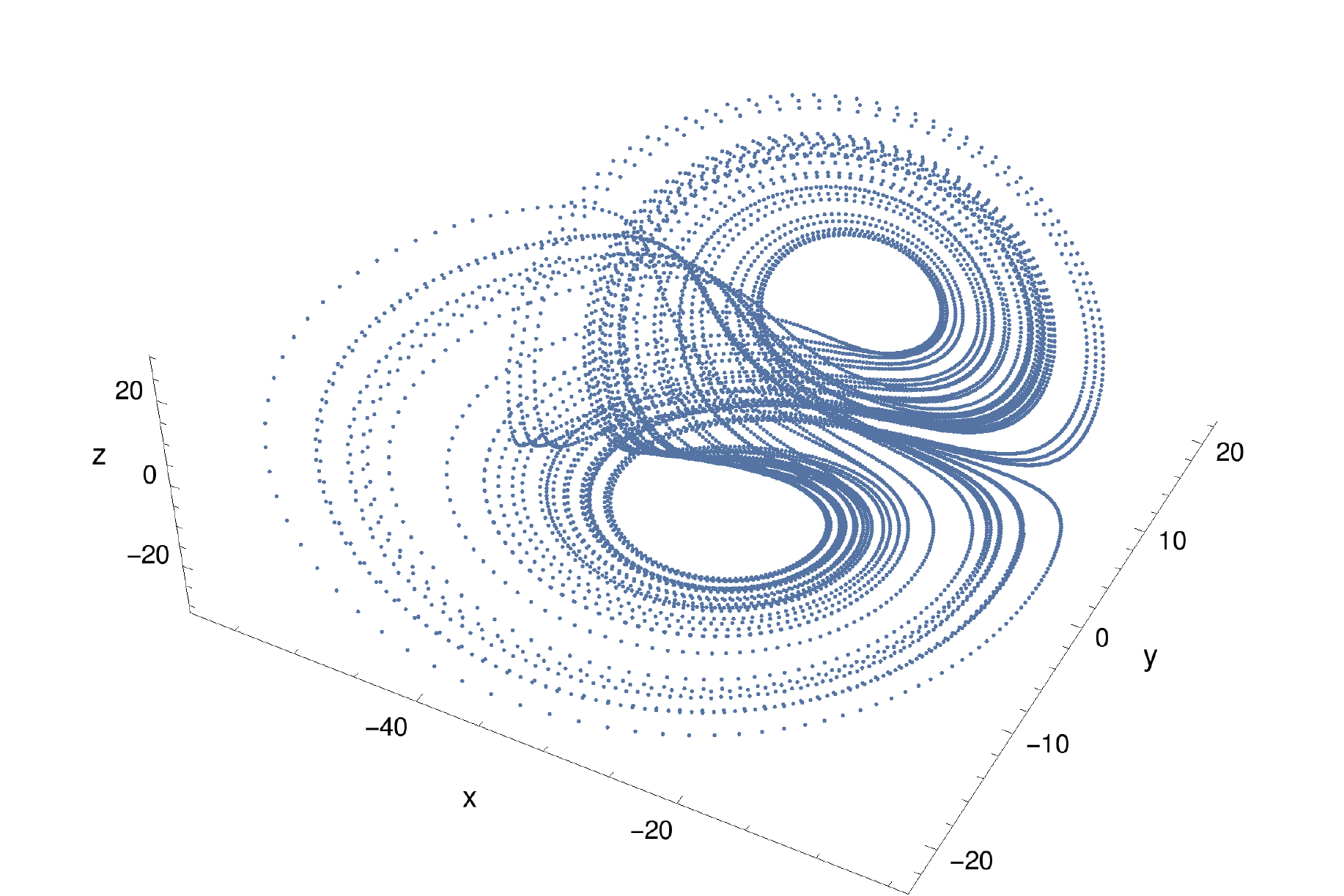}
	\caption{Chaotic behavior of BG system with $\alpha = 0.98$, $\mu = 10, \omega = -2.667, A = 26, b= 1$. Initial condition is $(x_0,y_0,z_0) = (4,0.5,0.1)$. }
	\label{fig:phaseBG98}
\end{figure}

Bhalekar and Daftardar-Gejji introduced a new chaotic system \cite{bhalekar2011new}, which is referred as BG system. 
The fractional version of BG system is given as follows.
\begin{align}
\begin{split}\label{BG1}
D^{\alpha} x_{r} &= \omega x_{r} - y_{r}^2, \\
D^{\alpha} y_{r} &= \mu (z-y), \\
D^{\alpha} z_{r} &= A y_{r} - b z_{r} + x_{r} y_{r},
\end{split}
\end{align} where $\omega <0, \mu >0$ and $A, b$ are parameters and $0 < \alpha \leq 1$.
BG system was analyzed by Deshpande and Daftardar-Gejji \cite{deshpande2017hopf} and shown to be chaotic for $\alpha_c < \alpha \leq 1$, where $\alpha_c$ denotes the critical value for $\alpha$ below which the system loses chaos. Moreover the authors have shown that $\alpha_c > \frac{2}{3}$. Figure (\ref{fig:phaseBG98}) shows phase portrait of Bhalekar-Gejji system for fractional order $\alpha = 0.98$ with parameter values $\mu = 10, \omega = -2.667, A = 26, b= 1$ with initial condition $(x_0,y_0,z_0) = (4,0.5,0.1)$. The LLE of the BG system under this case is calculated and found out to be LLE = $0.8806$. Thus the BG system shows chaotic behavior. 

In this section, we synchronize fractional order commensurate Vallis system \eqref{vallis1} as a Drive system with the fractional order BG system \eqref{BG1} as a Response system using active control functions $u_1(t), u_2(t), u_3(t)$.

Let $e_1 := x_{r} - x$, $e_2:= y_r - y$ and $e_3 : = z_r - z$ be the error variables. Then \eqref{vallis1} and \eqref{BG1} along with active control functions gives the error system  
\begin{align}\label{error1}
\begin{split}
D^{\alpha}e_1 &= (\omega-C) e_1 + B e_2 + (Cp - y_{r}^2 + \omega x + C x_r - B y_r) + u_1(t),\\
D^{\alpha}e_2 &= \mu e_3 - \mu e_2 + (\mu z - x z - (\mu-1) y) + u_2(t),\\
D^{\alpha}e_3 &= A e_2 - b e_3 + (x_r y_r + x y -1 + A y - (b-1) z ) + u_3(t). 
\end{split}
\end{align}
 
Let us choose $u_1, u_2, u_3$ as 
\begin{align}
\begin{split}\label{Us}
u_1 (t) &= -Cp + y_{r}^2 - \omega x - C x_r + B y_r + V_1(t), \\
u_2 (t) &= -\mu z + x z + (\mu-1) y + V_2(t), \\
u_3 (t) &= -x_r y_r - x y +1 - A y + (b-1) z + V_3(t), 
\end{split}
\end{align} where $V_1, V_2, V_3$ are linear functions of $e_1, e_2, e_3$, whose choice is not necessarily unique. Substituting \eqref{Us} into \eqref{error1} we get
\begin{align}\label{error2}
\begin{split}
D^{\alpha}e_1 &= (\omega-C) e_1 + B e_2 + V_1(t),\\
D^{\alpha}e_2 &= \mu e_3 - \mu e_2 + V_2(t),\\
D^{\alpha}e_3 &= A e_2 - b e_3 + V_3(t). 
\end{split}
\end{align}

The two systems \eqref{vallis1} and \eqref{BG1} will be synchronized if and only if the error system \eqref{error2} is asymptotically stable. We choose values of $V_1, V_2, V_3$ such that the eigenvalues $\lambda_i, ~ i=1,2,3$ of system \eqref{error2} satisfy $\mnorm{\arg(\lambda_i)} > \frac{\pi \alpha}{2}$. This will ensure asymptotic stability of the system \eqref{error2}.

In particular let
\begin{align}
\begin{split}\label{Vs}
V_1(t) &= - (\omega - C + K_1) e_1, \\
V_2 (t)&= (\mu - K_2) e_2, \\ 
V_3 (t) & = - A e_2 + (b-K_3) e_3,
\end{split} 
\end{align} where $K_1, K_2, K_3$ are positive real numbers called as control parameters. In view of \eqref{Vs}, system \eqref{error2} becomes
\begin{align}
D^{\alpha} \begin{pmatrix}
e_1 \\ e_2 \\ e_3
\end{pmatrix} = \begin{bmatrix}
-K_1 & B & 0 \\
0 & - K_2 & \mu \\
0 & 0 & - K_3
\end{bmatrix} \begin{pmatrix}
e_1 \\ e_2 \\ e_3
\end{pmatrix},
\end{align} having eigenvalues as $-K_1, -K_2, -K_3$. For positive values of $K_i$, the system is asymptotically stable. In particular, for $K_1 = K_2 = K_3  = 4$, system \eqref{error2} is asymptotically stable and system \eqref{vallis1} and \eqref{BG1} will synchronize. 

We simulate the Drive system \eqref{vallis1} for the parameter values $B = 150, C = 4, p = 0.35$ and initial condition $(5.86, 0.165, 0.028)$ and for the Response system \eqref{BG1}, we consider $\omega = -2.667, \mu = 10, A = 26, b = 1$ and initial condition $(4,0.5,0.1)$. Fractional order $\alpha$ is taken as $0.98$.

\begin{figure}[H]
	\centering
	\begin{subfigure}{0.7\linewidth}
		\includegraphics[width=\textwidth]{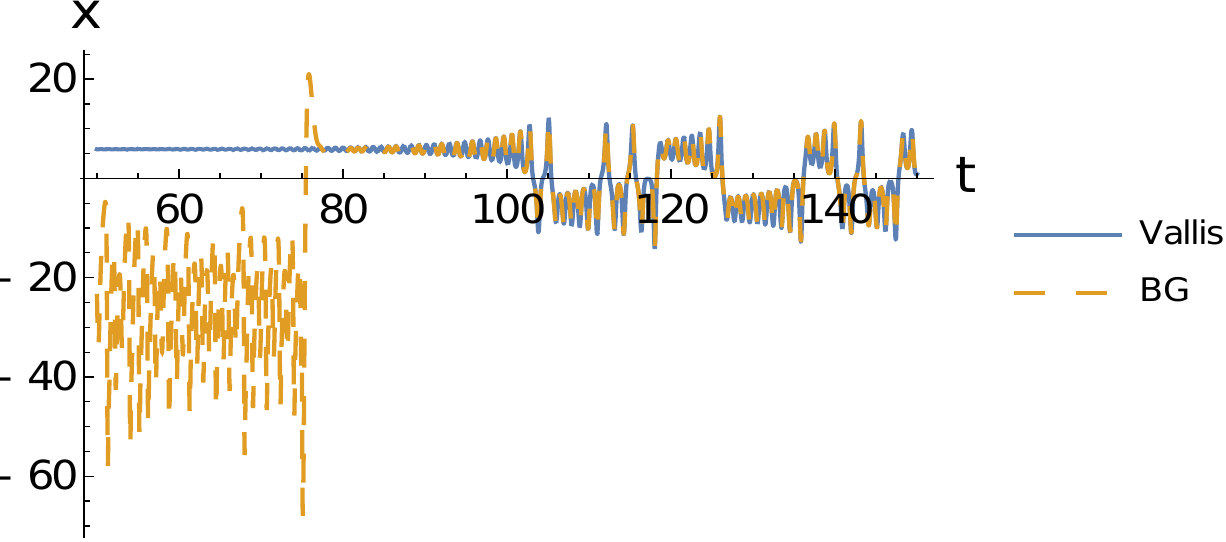}
	\end{subfigure}\\
	\begin{subfigure}{0.7\linewidth}
		\includegraphics[width=\textwidth]{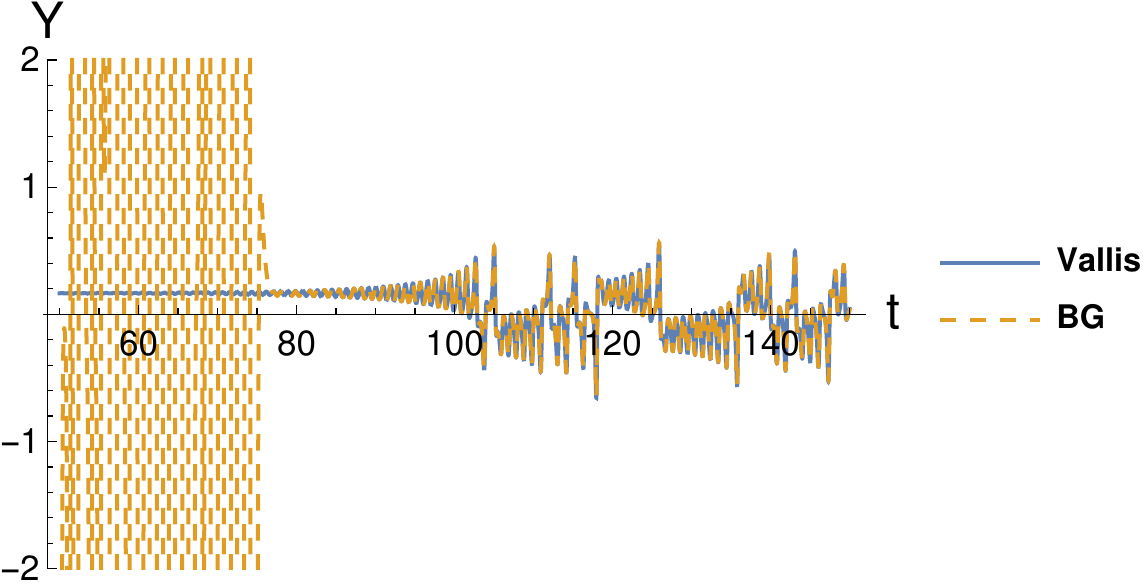}
	\end{subfigure}\\
	\begin{subfigure}{0.7\linewidth}
		\includegraphics[width=\textwidth]{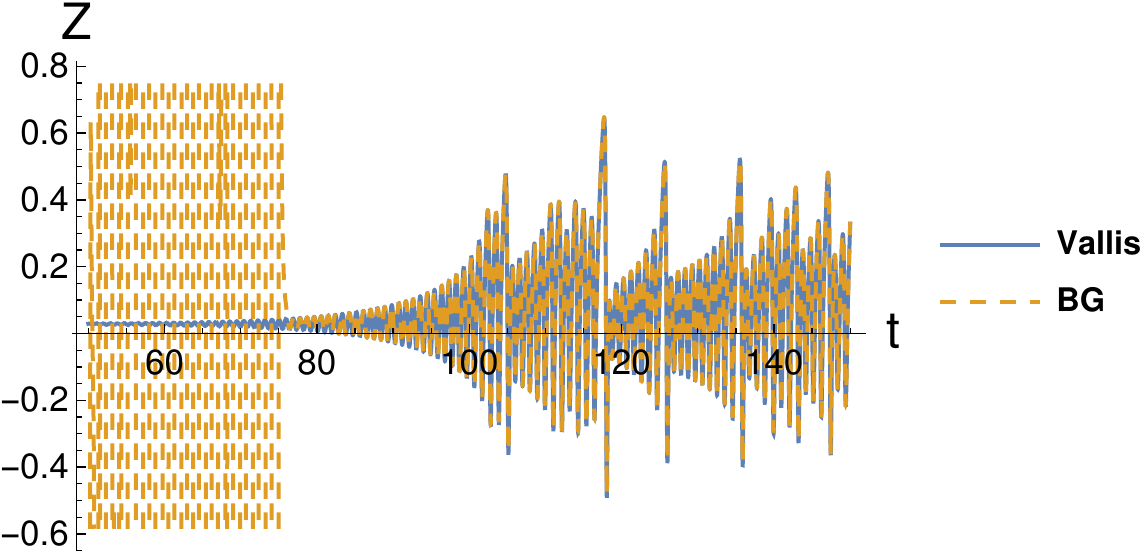}
	\end{subfigure}
	\caption{Evolution of time-series of $x, y, z$ variables of Vallis system \eqref{vallis1} (shown by Blue continuous line) and BG system \eqref{BG1} (Orange, dashed line) for $\alpha = 0.98$. The control is triggered at $t=75$.}
	\label{fig:synch1}
\end{figure}

\begin{figure}[H]
 	\begin{subfigure}{0.45\linewidth}
 		\includegraphics[width=\textwidth]{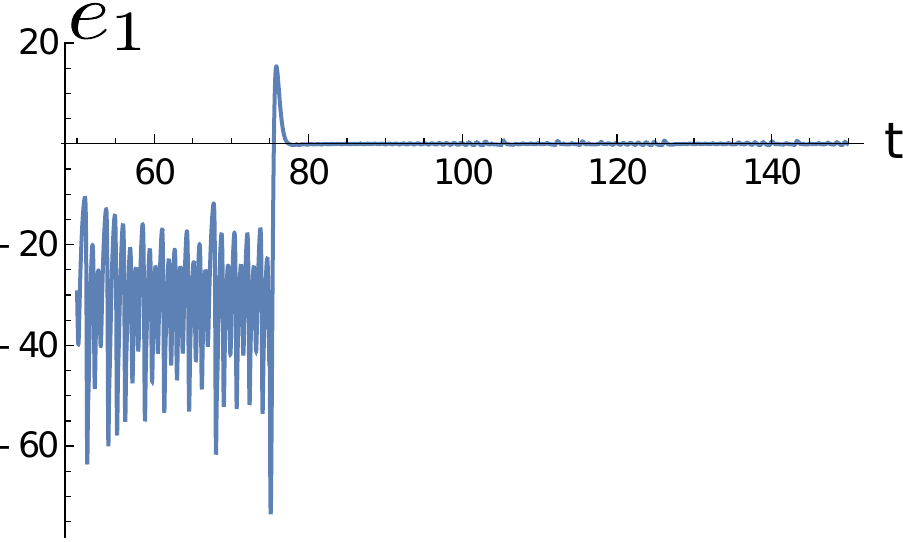}
 	\end{subfigure}
 	\begin{subfigure}{0.45\linewidth}
 		\includegraphics[width=\textwidth]{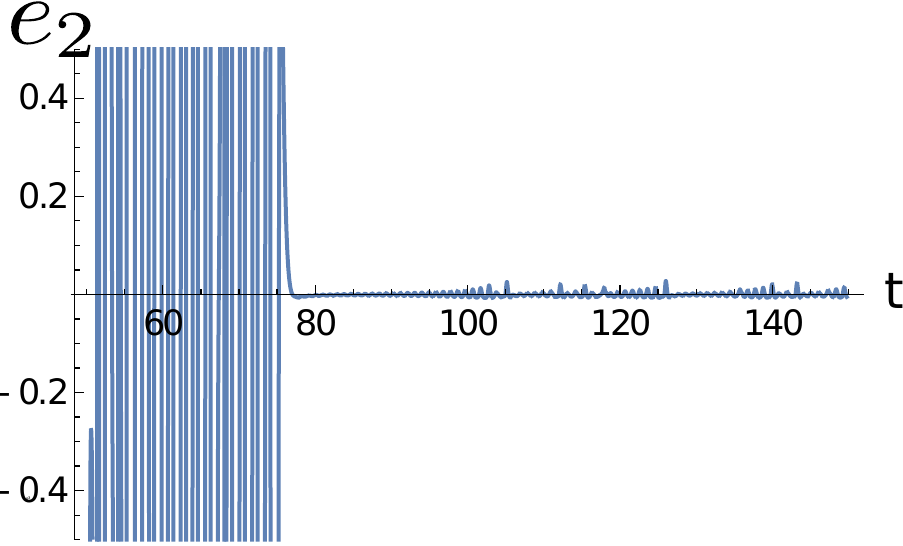}
 	\end{subfigure}\\
 	\begin{subfigure}{0.5\linewidth}
 		\includegraphics[width=\textwidth]{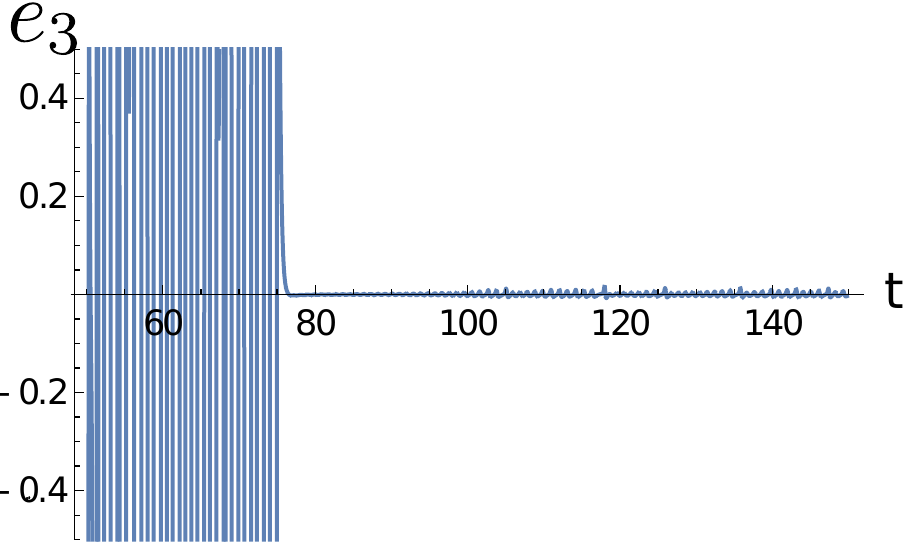}
 	\end{subfigure}
 	\caption{Evolution of error $e_1, e_2, e_3$ for $\alpha = 0.98$. The control is triggered at $t=75$.}
 	\label{fig:synch2}
 \end{figure}

Both the systems are allowed to evolve independently till $t=75$, after which the active control is switched on. Figure (\ref{fig:synch1}) shows timeline of $x ,y ,z$ variables of Vallis system (Blue) and BG sytem (Orange and dashed). It is clear that after $t=75$, both system synchronize with each other. Time evolution of corresponding error variables is shown in fig. (\ref{fig:synch2}).

\section{Conclusions} \label{conclusions}
Vallis proposed a model for El-Ni\~{n}o phenomenon and showed that El-Ni\~{n}o effect is related to the chaotic behavior of the Vallis system. Hence it is important to study the chaos in Vallis system.

 In the present article, fractional version of commensurate as well as incommensurate Vallis system has been studied in detail. The effect of variation of system parameter $p$ and the fractional order, on the chaotic behavior of the Vallis system is investigated using  bifurcation diagrams and largest Lyapunov exponents. In particular it is observed that the range of the parameter $p$ for which the system is chaotic; reduces rapidly as the fractional order is reduced. For $\alpha \leq 0.965$, the system is no longer chaotic for any value of parameter $p$. 
 
 The critical fractional order for which fractional Vallis system is no longer chaotic has been computed in case of $B=105,~ C = 4,$ and $p = 0.35$. This critical order is found to be $0.975 \;\times\;3 \;= 2.925$ for the commensurate Vallis system,  while the lowest critical order for the incommensurate fractional system, is found to be  $2.71$. 
 Further we have synchronized fractional Vallis system with fractional Bhalekar-Gejji sytem.      
 
 \section*{Acknowledgments}
 Authors thank the Center for Development of Advanced Computing (CDAC), Pune for providing National Param Supercomputing Facility. 
  

\end{document}